# Curved graphene nanoribbons derived from tetrahydropyrene-based polyphenylenes via one-pot K-region oxidation and Scholl cyclization

Sebastian Obermann,[a] Wenhao Zheng,[b] Jason Melidonie,[a] Steffen Böckmann,[c] Silvio Osella,[d] Lenin Andrés Guerrero León,[a] Felix Hennersdorf,[e] David Beljonne,[f] Jan J. Weigand,[e] Mischa Bonn,[b] Michael Ryan Hansen,[c] Hai I. Wang,[b] Ji Ma,[a,g,*] Xinliang Feng[a,g,*]

Precise synthesis of graphene nanoribbons (GNRs) is of great interest to chemists and materials scientists because of their unique opto-electronic properties and potential applications in carbon-based nanoelectronics and spintronics. In addition to the tunable edge structure and width, introducing curvature in GNRs is a powerful structural feature for their chemi-physical property modification. Here, we report an efficient solution synthesis of the first pyrene-based GNR (**PyGNR**) with curved geometry via one-pot K-region oxidation and Scholl cyclization of its corresponding well-soluble tetrahydropyrene-based polyphenylene precursor. The efficient A₂B₂-type Suzuki polymerization and subsequent Scholl reaction furnishes up to ~35 nm long curved GNRs bearing cove- and armchair-edges. The construction of model compound (**1**), as a cutout of **PyGNR**, from a tetrahydropyrene-based oligophenylene precursor proves the concept and efficiency of the one-pot K-region oxidation and Scholl cyclization, which is clearly revealed by single crystal X-ray diffraction analysis. The structure and optical properties of **PyGNR** are investigated by Raman, FT-IR, solid-state NMR and UV-Vis analysis with the support of DFT calculations. **PyGNR** shows the absorption maximum at 680 nm, exhibiting a narrow optical bandgap of ~1.4 eV, qualifying as a low-bandgap GNR. Moreover, THz spectroscopy on **PyGNR** estimates its macroscopic charge mobility $\mu$ as ~3.6 cm²·V⁻¹·s⁻¹, outperforming other curved GNRs reported via conventional Scholl reaction.

## Introduction

Structurally well-defined graphene nanoribbons (GNR) are a constantly developing field, amassing more and more attention due to their unique opto-electronic properties[1–3] and potential to be integrated into future nanoelectronics and spintronics.[3–9] The physicochemical properties of atomically precise GNRs are highly dependent on their edge structures and widths.[6,7,10–13] In the well-investigated armchair-edged GNRs (AGNR), three different families of ribbons can be distinguished as a function of carbon atoms lateral to the ribbon propagation direction, exhibiting semiconducting properties.[6,14–21] On the other hand, zigzag-edged GNRs (ZGNR) possess edge states and metallic character, but could only be synthesized on surface to date, drastically limiting their applicability.[2,6,22–24] In stark contrast to the destructive and largely uncontrolled top-down methods, the mentioned examples were achieved by bottom-up synthesis, an approach also rendering mixed edges possible. For instance, mixing armchair edge with zigzag edge can create intriguing physical phenomena, as recently exemplified by the topological quantum phase in GNRs.[7,25–27] Aside from the edge type, curvature (or non-planarity) is another powerful structural feature for tuning the properties of GNRs,[28,29] such as liquid-phase dispersibility and charge transport behavior.[30–32] A common synthetic pathway to achieve curvature in GNRs is to introduce the structural motif of small [n]helicene units on its periphery, such as [4]helicene or [5]helicene,[33–35] creating the cove edge [12,36–38] or fjord edge[39] respectively. For example, the regular incorporation of cove subunits by elimination of a single carbon atom along the opposite ZGNR edges can produce a series of curved GNRs with periodic cove-zigzag edge structure, providing GNRs with well-tunable band structures and effective masses (Figure 1a).[11] In the context of this structural modification of ZGNRs towards cove-edge topologies, it raises the synthetic question if cove-edges can also be accessed from the parent design of AGNR (Figure 1b).

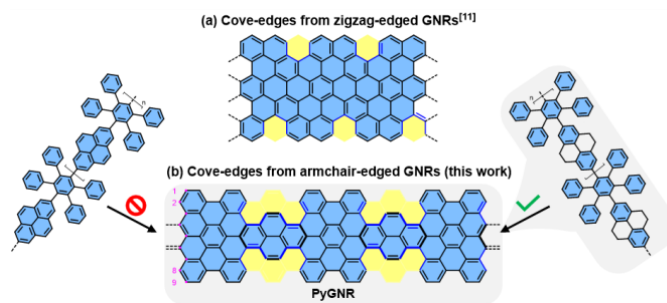

**Figure 1** (a) One example of curved GNRs with periodic cove-zigzag edges; (b) Curved GNRs obtained by incorporation of cove units along the 9-AGNR edges, reported in this work.

Pyrene (**2**) is one of the widely used building blocks in material science.[40] Its biphenyl-like structure with two double bond characteristics, called the K-region (Scheme 1), distinguishes pyrene from other polycyclic aromatic hydrocarbons (PAHs).[40,41] Moreover, its full-zigzag edge structure renders it a versatile motif for constructing highly conjugated π-systems with tailored edge topology and functions. Although it is possible to functionalize pyrene on every position by various synthetic methods,[42–44] its incorporation into large PAHs (or nanographenes) via Scholl reaction of pyrene-based dendritic oligophenylene precursors often causes problems due to its high reactivity and rich electron density.[45,46] Therefore, the expansion of pyrene-based precursors towards the synthesis of GNRs remains a synthetic challenge.

Herein, we demonstrate an efficient solution synthesis of a novel curved pyrene-based GNR (**PyGNR**) bearing cove- and armchair-edges from the predesigned tetrahydropyrene-based polyphenylene precursor (**P1**). It is worth mentioning that, when trying to obtain **PyGNR** from pyrene-embedded polyphenylene directly (Figure 1b), only short and insoluble oligomers could be obtained. To overcome this obstacle, the unsaturated polymer **P1** is first obtained by $A_2B_2$-Suzuki polymerization of 2,7-bis(4,4,5,5-tetramethyl-1,3,2-dioxaborolan-2-yl)-4,5,9,10-tetrahydro-pyrene (**4**) and alkyl-chain equipped tetraphenyl-substituted 1,4-diiodobenzene (**12**). Subsequently, one-pot oxidation of the tetrahydropyrene K-regions towards pyrene moieties together with the Scholl reaction enables the desired formation of **PyGNR**. The resultant **PyGNR** reaches a length of up to ~35 nm according to the gel-permeation chromatography (GPC) analysis of **P1**. As a cutout of **PyGNR**, model compound **1** was successfully synthesized to evaluate the efficiency and feasibility of the one-pot cyclodehydrogenation process. The structure and optical properties of **PyGNR** were well investigated by solid-state NMR, FT-IR, Raman and UV-Vis spectroscopy. Due to the unique edge structure, a narrow optical bandgap of ~1.4 eV was determined according to the UV-Vis spectra, slightly below the value of 1.68 eV predicted by DFT. Furthermore, time-resolved terahertz spectroscopy reveals a macroscopic charge mobility of ~3.6 $cm^2V^{-1}s^{-1}$, outperforming several other examples[37,39] of curved GNRs obtained by conventional Scholl reactions.

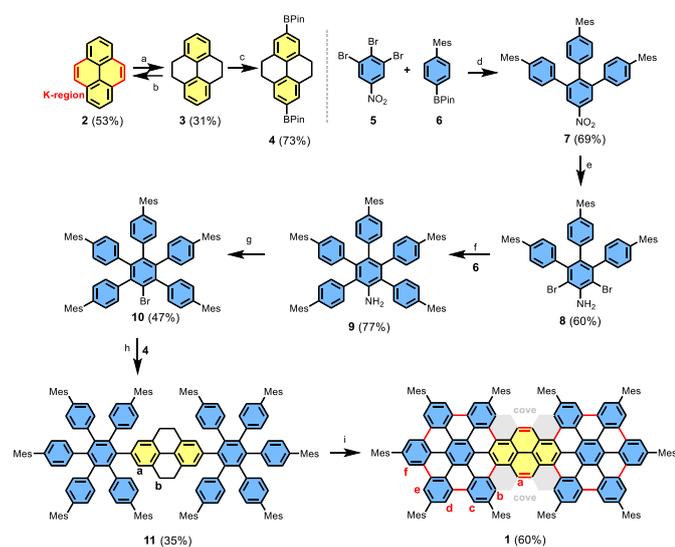

**Scheme 1** Reoxidation of 4,5,9,10-tetrahydropyrene (**3**) and synthetic route to model compound **1**. Reagents and conditions: a) i. Raney-Nickel, EtOAc, rt, 48 h, ii. Pd/C (10%), EtOAc, $H_2$, 60 °C, 24 h; b) $FeCl_3$ (10 eq.), $MeNO_2$, $CH_2Cl_2$, rt, 5 min; c) $(BPin)_2$, 4,4'-di-tertbutyl-2,2'-bipyridine, $[Ir(OMe)(COD)]_2$, THF, 80 °C, overnight; d) **5**, **6**, $Pd(dppf)Cl_2 \cdot CH_2Cl_2$, $K_3PO_4$, PhMe/$H_2O$, 115 °C, overnight; e) i. $SnCl_2$, EtOH/EtOAc, reflux, overnight, ii. NBS, $CHCl_3$, rt, 10 min; f) **6**, $Pd(dppf)Cl_2$, $K_3PO_4$, PhMe/$H_2O$, 115 °C, 48 h; g) $CHBr_3$, isoamylnitrite, 100 °C, overnight; h) $Pd(PPh_3)_4$, $K_2CO_3$, PhMe/$H_2O$, 115 °C, 48h; i) DDQ, $MeSO_3H$, $CH_2Cl_2$, rt, 30 min, then DDQ, TfOH, rt, 3 min.

## Results and Discussion

**Synthesis of model compound 1**

The synthesis of model compound **1** is illustrated in Scheme 1. Considering the solubility issue of **1**, mesitylene side groups were installed instead of the more common *tert*-butyl groups. First, 4,5,9,10-tetrahydropyrene (**3**) was obtained from the commercially available pyrene (**2**) after desulfurization with Raney-Nickel and hydrogenation under pressure in the presence of Pd/C among over-hydrogenated products in 31% yield (Scheme 1). Then, Iridium-catalyzed borylation[42] was used to furnish compound **4** in 73% yield. Slightly modifying the synthesis route of Lungerich *et al.*[47] allowed us to synthesize compound **10** starting from 1,2,3-tribromo-5-nitrobenzene (**5**, see ESI). A Suzuki coupling reaction between **5** and **6** afforded compound **7** in 69% yield, followed by a two-step sequence of reduction and bromination to furnish compound **8** in 60% yield. Another Suzuki coupling of **8** and **6** gave compound **9** in 77% yield. In a Sandmeyer-type reaction, aniline **9** was brominated to compound **10** with a yield of 47%. Finally, building blocks **4** and **10** were cross-coupled by the Suzuki reaction, providing the oligophenylene precursor **11** in a low yield of 35% due to the steric hindrance of the inner mesitylene units in **11**. For the cyclodehydrogenation of **11**, we initially examined the classic Scholl conditions by use of $FeCl_3$ or triflic acid with DDQ in $CH_2Cl_2$, but failed to obtain a defined product. We then turned to a milder approach using methylsulfonic acid and DDQ, which furnished a time-stable MALDI-TOF signal 12 proton masses above the desired signal, indicating a partially closed intermediate (ESI Figure S31).

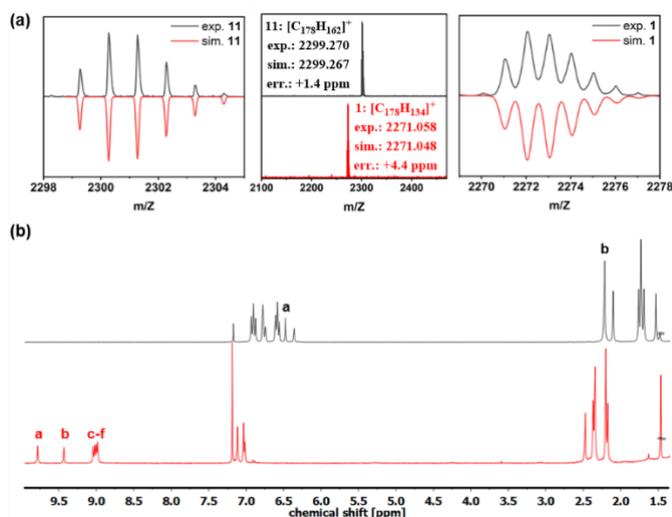

**Figure 2** (a) MALDI-TOF-MS of compounds **11** and **1**. (b) $^1$H-NMR spectra of compound **11** (top) and compound **1** (bottom), recorded in CDCl$_3$ at 300 MHz (proton labels see Scheme 1).

After adding the stronger triflic acid to the reaction mixture, the bright red compound **1** was obtained in 60% yield after only three minutes of additional reaction time.

The chemical identity of **1** was confirmed by high-resolution mass spectroscopy and NMR spectroscopy, whose key changes for compounds **11** and **1** are illustrated in Figure 2, respectively. The MALDI-TOF spectra show a mass difference of 28 proton masses, meaning that in addition to the cyclodehydrogenation of 24 protons, 4 additional hydrogens were removed from the tetrahydropyrene moiety to reestablish the pyrene K-region. In addition to the reduction of NMR signals in the case of compound **1**, the total amount of protons in the aliphatic region was reduced from 98 to 90 (ESI Figure S18/S20), further confirming the successful conversion of the tetrahydropyrene moieties. The anti π-π-stacking properties of the mesitylene side groups in **1** enabled us to assign the protons using two-dimensional COSY- and NOESY-NMR spectroscopy (ESI Figure S21/S22). Furthermore, a shift in the downfield due to the aromatic ring currents was observed, especially for the protons *a* and *b* (proton labels seen in Scheme 1) resulting from their location on the cove edge.

**Crystallographic analysis of compounds 11 and 1**

Crystals suitable for X-ray crystallographic analysis were obtained for **11** and **1**, respectively (Figure 3). In the case of **11**, a suitable crystal was grown from **11** dissolved in CH$_2$Cl$_2$ by slow evaporation of the solvent. Single crystal X-ray diffraction analysis (SCXRD) reveals an unconjugated phenylic structure combined with twisted sp$^3$-bonds of the central tetrahydropyrene unit. The sp$^3$-carbon single bonds are revealed by the bond lengths of 1.57 Å and 1.58 Å, respectively. Compound **11** crystallizes in the triclinic space group P1. In the case of **1**, a single crystal was obtained by slow diffusion of methanol in its CS$_2$-solution. The SCXRD analysis reveals **1** as a fully conjugated, curved polycyclic hydrocarbon with four cove edges. The hexa-*peri*-hexabenzocoronene (HBC) units have been fully fused and share a pyrene unit in the center, revealing the successful conversion of the tetrahydropyrene moieties back to a fully sp$^2$-hybridized pyrene core and supporting the proposed one-pot K-region oxidation and Scholl cyclization. The cove edges between the pyrene K-region and the HBC moieties possess an average torsion angle of 21°, resulting in curvature of the molecule. In contrast to compound **11**, compound **1** crystallizes in the monoclinic space group P2$_1$/c. The bond lengths in the K-regions of compound **1** are significantly shorter, about 1.35 Å (1.337 Å and 1.361 Å, respectively), matching with the value for pyrene (**2**) reported by the literature and indicating double bond character.[48] In the solid packing, the crystal layers of **1** are slightly slipped with the π-surfaces facing towards each other, forming π-π-stacks (Figure 3c).

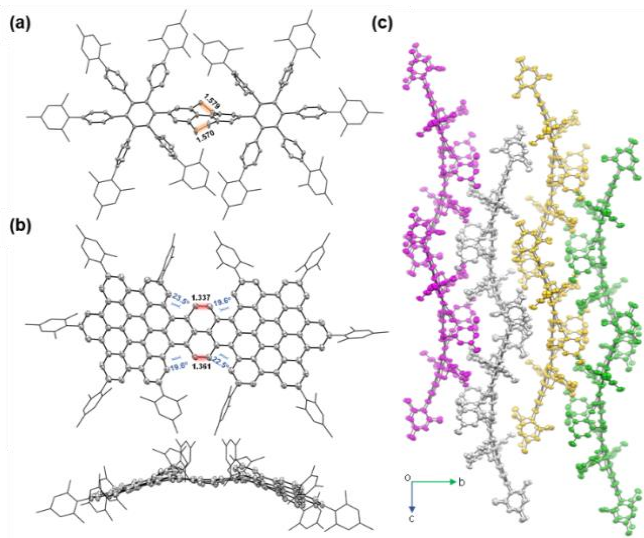

**Figure 3** X-ray crystallographic structure (ORTEP drawing at the 50 % probability level) of (a) **11** and (b) **1** (top and side view). (c) Crystal-packing of **1**. Hydrogen atoms and solvent molecules have been omitted for clarity.

**Synthesis of PyGNR**

After confirming the model reaction formed the desired compound **1**, we further extended the synthesis towards **PyGNR** based on tetrahydropyrene monomer **4**. Early attempts of this A$_2$B$_2$ Suzuki polymerization with 2,7-diborylated pyrene gave only oligomers according to linear mode MALDI-TOF MS measurements, which remained insoluble and therefore unavailable for further solution processing (ESI Figure S33). When these oligomers were exposed to Scholl-reaction conditions as a suspension, partial cyclization and decomposition were observed. In consequence, the 4,5,9,10-tetrahydropyrene containing polymer **P1** (Scheme 2) was designed. The monomer **12**,[15] substituted with 3,7-dimethyloctyl chains, was chosen as the coupling partner for the diborylated pyrene derivative **4** to increase the processibility of the GNR later. In contrast to pyrene (**2**), when switching to the tetrahydropyrene derivative **4**, soluble polymers were obtained when refluxing the monomers, Pd(PPh$_3$)$_4$, K$_2$CO$_3$ in toluene and H$_2$O for 48 hours. Linear mode MALDI-TOF MS of the crude material revealed groups of signals up to *m/Z* ~ 10 000, spaced by the repetition unit of 1145 g·mol$^{-1}$ (ESI Figure S32).

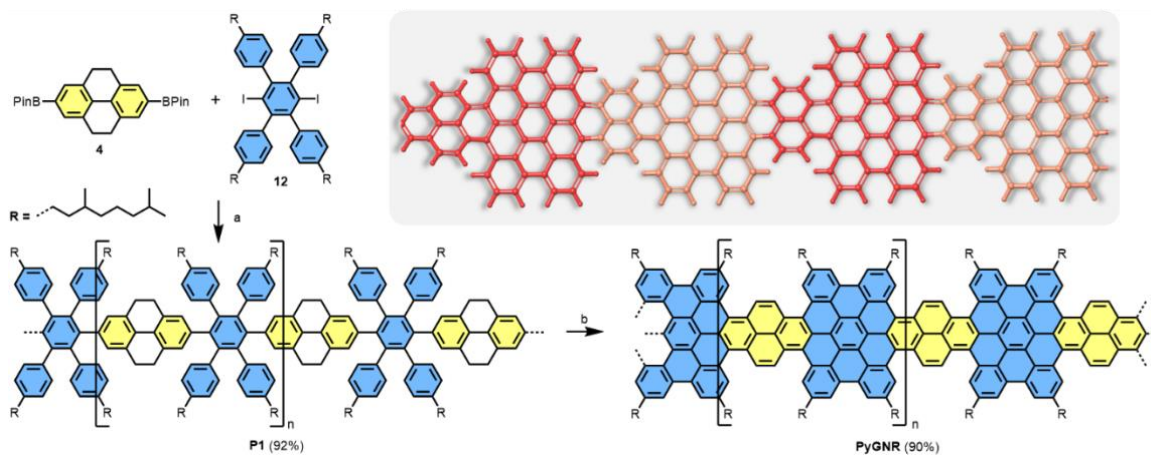

**Scheme 2** Synthetic route to **PyGNR**. Reagents and conditions: a) **4** (stoichiometric), **12**, Pd(PPh$_3$)$_4$, K$_2$CO$_3$, Aliquat 336, PhMe/H$_2$O, 110°C, 48 h; b) FeCl$_3$, MeNO$_2$, CH$_2$Cl$_2$, rt, 7 d. Inset is the optimized geometry of **PyGNR**, aliphatic side chains omitted for clarity.

After washing with HCl/MeOH, filtration and subsequent dissolution and filtration in chloroform, the crude polymer **P1** was obtained after removing the chloroform *in vacuo*. **P1** was then divided into 4 fractions by recycling GPC and further analyzed by analytical GPC against polystyrene standards, revealing a number average molar mass of $M_n$ ~ 24 000 Da and a narrow polydispersity index (PDI) of 1.2 for the heaviest (10 wt%) crude polymers (ESI Figure S35-37), significantly enhancing the molecular weight compared to other GNRs reported via the Suzuki polymerization method.[15,39] The protons could be assigned to their positions in **P1** by $^1$H-NMR spectroscopy (ESI Figure S29). Due to **P1**'s excellent solubility, even $^{13}$C spectra could be recorded effortlessly (ESI Figure S30). Finally, the Scholl reaction of the first fraction of **P1** with FeCl$_3$ (7 equiv/H) as the Lewis acid and oxidant in CH$_2$Cl$_2$ for 7 days yielded the **PyGNR**. The resultant **PyGNR** possesses combined cove and armchair edges with an estimated average length of ~35 nm based on the $M_n$ of the corresponding **P1**.

**Structural characterizations of PyGNR**

The structures of precursor **P1** and **PyGNR** were then characterized by a combination of solid-state $^1$H and $^{13}$C{$^1$H} MAS NMR ($^1$H decoupled $^{13}$C magic angle spinning NMR) at a high spinning speed of 62.5 kHz (Figure 4a,b), FT-IR and Raman (Figure 4c,d). While the precursor **P1** shows well-defined signals in the $^1$H MAS NMR spectrum (Figure 4a) for both the aromatic and aliphatic protons centered around 0.9 ppm and 6.6 ppm, respectively, broader lines are observed for **PyGNR** along with a downfield shift of the aromatic protons to 8.9 ppm. This can be attributed to a more rigid structure and extended π-conjugation of **PyGNR** compared to **P1**. Deconvolution of the quantitative $^1$H MAS NMR spectra shows a lower relative content of aromatic protons in **PyGNR** compared to **P1** as a result of the graphitization and matches the expected relative intensities within the expected error margins. The graphitization process can also be confirmed by the $^{13}$C{$^1$H} MAS NMR spectra that were acquired with a short recycle delay of 3.5 s (Figure 4b). Under the chosen experimental conditions the $^{13}$C signal of the side chains only shows low intensity. In addition, the structural evolution from **P1** to **PyGNR** is also supported by 2D solid-state NMR experiments (see ESI chapter 9).

The FT-IR spectra of **P1** (Figure 4c) exhibit typical vibrational modes for its structural elements. A more detailed discussion can be found in the ESI, chapter 10. To assist in the identification of the corresponding vibrations, DFT calculations on the HSE/6-31G(d) level of theory were performed. The peak at 885 cm$^{-1}$, which was identified as the SOLO-mode (CH-wagging of an aromatic proton without adjacent protons, orange)[49] of the tetrahydropyrene moiety, confirms once more the presence of unsaturated bonds in **P1**. The FT-IR spectrum of **PyGNR** was then compared to **P1**. The abundant aromatic CH-stretchings in the range of 3060-3020 cm$^{-1}$ were diminished in the graphitized ribbon due to the cyclodehydrogenations. The CH-stretchings from the aliphatic side chains from 3000-2750 cm$^{-1}$ were retained. Further, characteristic vibrations were investigated in the fingerprint region and identified by comparison to the DFT prediction. First, the SOLO- modes can be found around 882 cm$^{-1}$ (pink), a further indication of successful graphitization. Interestingly, **PyGNR**s CH-wagging of the cove SOLO-mode and pyrene DUO moieties (CH-wagging of two adjacent aromatic protons) can be found in a joint vibration mode around 840 cm$^{-1}$ (red), demonstrating the successful conversion of the tetrahydro-moieties into pyrene K-regions. This claim is further supported by the ring vibrations of the pyrene K-region, which can be attributed to the newly emerged signals around 790-770 cm$^{-1}$ (blue), absent in **P1**. Last, the broad signal at 640 cm$^{-1}$ (brown) gives another strong indication of successful graphitization, as the DFT-prediction assigns these modes as in plane aromatic CC-scaffold vibrations. The Raman spectrum of **PyGNR** (Figure 4d) shows two strong D and G bands at ~1340 and ~1590 cm$^{-1}$. Furthermore, double resonance peaks can be found at ~2605, ~2880 and ~3170 cm$^{-1}$, which can be assigned as the 2D, D+G and 2G bands. The peak positions are in line with other bottom-up solution-synthesized GNRs reported in the literature.[13,21,37]

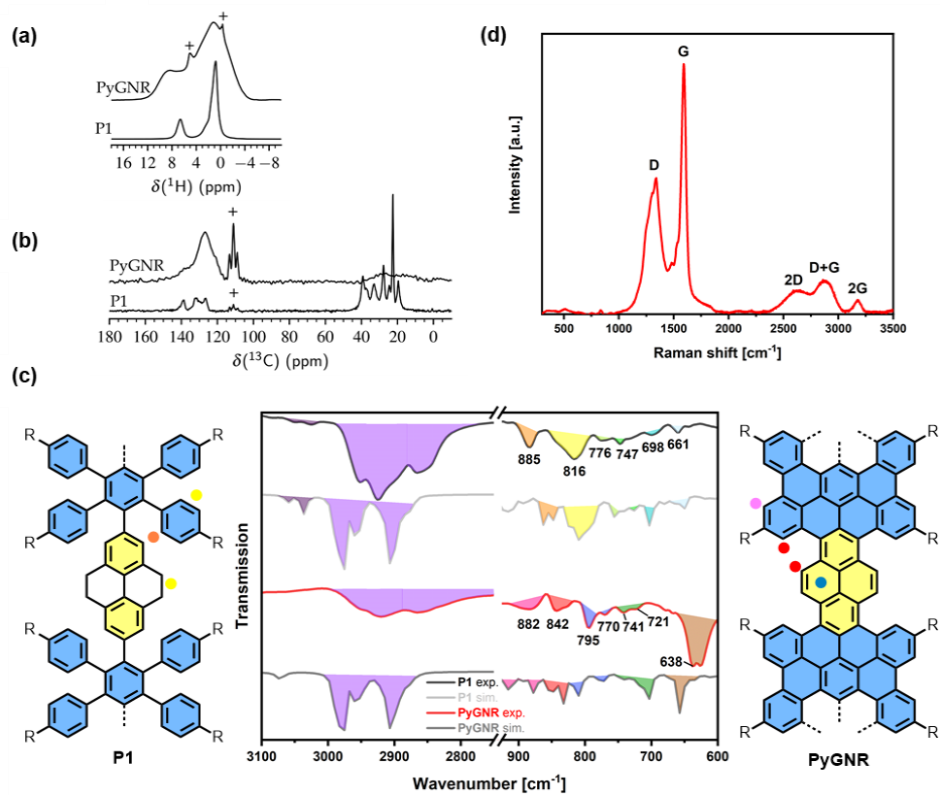

**Figure 4** (a) Quantitative $^1$H MAS NMR spectra of **P1** and **PyGNR** with 64 s recycle delay. Signals of remaining solvent are marked with +; (b) $^{13}$C{$^1$H} MAS NMR spectra with a short recycle delay of 3.5 s and 25 kHz of low-power CW decoupling. The signal of PTFE spacer material is marked with +. All spectra were acquired at a static field of 11.74 T and a spinning frequency of 62.5 kHz. (c) FT-IR spectra of **P1** and **PyGNR**, as well as their predicted spectra using the HSE/6-31G(d) DFT level of theory. The fingerprint region (900-600 cm$^{-1}$) of the simulated spectra were scaled by factor 10; (d) Raman spectra of **PyGNR**, recorded at 514 nm.

**Optical property and THz spectroscopy study of PyGNR**

The UV-Vis absorption spectra were recorded in CHCl$_3$ solutions. An NMP dispersion of **PyGNR** that was allowed to settle overnight was used (Figure 5a). Due to its lack of conjugation, polymer **P1** (dark grey) possesses only high energy absorption bands starting with a first peak at 338 nm, which agrees well with the polymer fragment **11** (light gray), exhibiting peaks at similar wavelengths (326 nm and 267 nm). Significantly red-shifted to 559 nm due to its full conjugation, the first absorption of **1** (blue) can be found. Two distinct absorption peaks of **1** are visible at 437 nm and 367 nm. Compared to its short fragment **1**, the absorption spectrum of **PyGNR** (purple) is further red-shifted and displays an absorption peak with a maximum at ~680 nm as well as a shoulder at 485 nm, showing strong proof for graphitization compared to the high energy absorptions of **P1**. The optical bandgap was determined as 1.4 eV by a Tauc-plot, qualifying as a low bandgap GNR. The experimental result is in good agreement with the DFT-prediction of 1.75 eV (ESI Figure S2). In addition, the simulated TD-DFT curves for **1** (light blue) and **PyGNR** (light purple) are in excellent agreement l with the experimental spectra.

To investigate the charge transport properties of **PyGNR**, we conducted optical pump-THz probe (OPTP) spectroscopy measurements (see the experimental details in the ESI chapter 8).[50,51] Figure 5b shows the time-resolved complex photoconductivity of **PyGNR** dispersed in 1,2,4-trichlorobenzene, following above-gap optical excitation (here, 1.55 eV laser pulses). The rapid sub-picosecond rise of the real photoconductivity indicates charge carrier generation, and the following rapid decay can be assigned to the carrier trapping and/or the formation of bound electron-hole states (*i.e.*, excitons) due to the strong quantum confinement and the dielectric screening effect.[51,52] The exciton formation is featured by a large imaginary conductivity and a small real conductivity at the later time scale. To further study the free charge carriers' transport properties, we measured the conductivity spectrum around the peak photoconductivity (~1.2 ps after photoexcitation), as shown in Figure 5c (for experimental details see ESI chapter 8). The phenomenological Drude–Smith (DS) model was applied to describe the frequency-resolved complex conductivity (for details, see ESI).[53,54] The DS model

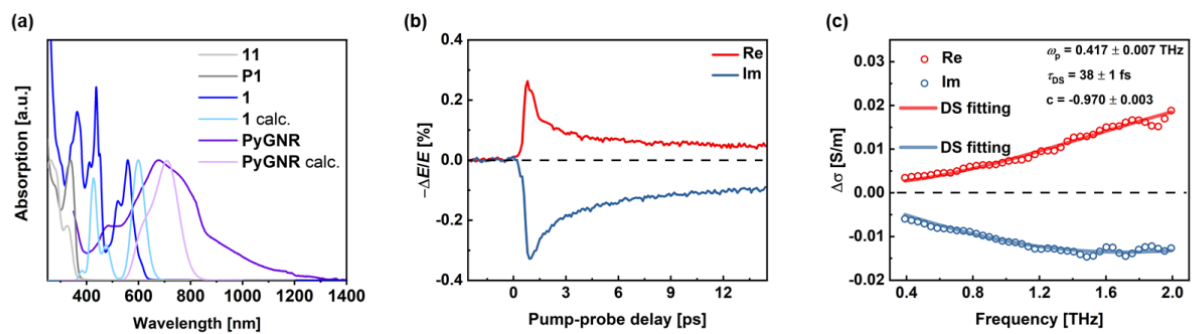

**Figure 5** (a) Absorption spectra of **1**, **11**, **P1** (CHCl$_3$), and **PyGNR** (NMP) as well as DFT-predicted **1** and **PyGNR**; (b) Time-resolved terahertz photoconductivity of **PyGNR** following photoexcitation. The absorbed photon fluence is ~1 mJ/cm$^2$. (c) The frequency-resolved THz complex conductivity measured ~1.2 ps after photoexcitation. The solid lines are fits to the Drude–Smith model.

considers the charge scattering occurring via preferential backscattering due to conjugation or structural deformation in materials. A parameter $c$ is introduced to characterize the backscattering probability, ranging from 0 (isotropic scattering) to −1 (100% backscattering). From the results we obtained an effective charge scattering time $\tau$ = 38 ± 1 fs and $c$ = −0.970 ± 0.003. Furthermore, by considering a reduced charge carrier mass m$^*$ = 0.50 m$_0$ from DFT (considering contributions from both electrons and holes; see the ESI for details), we estimate the macroscopic charge mobility $\mu\ (= \frac{e\tau}{m^*}(1+c))$ to be ~3.6 cm$^2$·V$^{-1}$·s$^{-1}$, outperforming similar examples of curved GNRs in the recent literature such as a a hybrid cove- and gulf-edged GNR,[37] and a 9AGNR based fjord-edged GNR.[29,39]

## Conclusion

In summary, we demonstrate the efficient synthesis of the first curved pyrene-based GNR (**PyGNR**) by introduction of cove units along the 9-AGNR edges. The synthetic strategy to obtain **PyGNR** was carried out in a previously unreported manner, regenerating the K-regions of the pyrene moieties in one step during the cyclodehydrogenation of a well-soluble unsaturated polyphenylene precursor **P1**. Proving the structural concept of the ribbon and the proposed one-pot oxidation, model compound **1** was also synthesized. The obtained **PyGNR** was fully characterized by solid-state NMR, FT-IR, Raman and UV-Vis spectroscopy, also supported by the DFT spectra simulations. THz study of **PyGNR** revealed a macroscopic charge mobility of ~3.6 cm$^2$V$^{-1}$s$^{-1}$, outperforming similar GNRs reported by conventional Scholl-cyclizations.[37,39] Furthermore, **PyGNR** is the first GNR with accessible pyrene K-regions, rendering further functionalization, e.g., by oxidation to diketones, possible.[43] Strategic use of tetrahydro-derivatives demonstrates the feasibility of obtaining novel GNRs in one step from sp$^3$-C containing polymer precursors and opens up possibilities for other GNRs suffering from polymer solubility issues or dehydrogenation problems.

## Conflicts of interest

The authors declare no conflict of interest.

## Acknowledgements

This research was financially supported by the EU Graphene Flagship (Graphene Core 3, 881603), ERC Consolidator Grant (T2DCP, 819698), H2020-MSCA-ITN (ULTIMATE, No. 813036), the Center for Advancing Electronics Dresden (cfaed), H2020-EU.1.2.2.- FET Proactive Grant (LIGHT-CAP, 101017821), the DFG-SNSF Joint Switzerland-German Research Project (EnhanTopo, No. 429265950). The authors gratefully acknowledge the GWK support for funding this project by providing computing time through the Center for Information Services and HPC (ZIH) at TU Dresden. S.O. thanks the Polish National Science Centre for funding (grant no. UMO-2020-39-I-ST4-01446). The computation was carried out with the support of the Interdisciplinary Center for Mathematical and Computational Modeling at the University of Warsaw (ICM UW) under grants no. G83-28 and GB80-24).The authors express their gratitude to Markus Göbel for the Raman measurement, Frank Drescher for the ESI-measurements, Enrique Caldera Cruz for the analytical GPC measurements and Tobias Nickel for the MALDI-TOF measurements.